\documentclass[12pt]{article}
 \usepackage{graphicx,amsmath,amssymb}
\usepackage{indentfirst}
 \usepackage[usenames]{color}
 \usepackage[colorlinks=true, urlcolor=navyblue, linkcolor=navyblue, citecolor=navyblue]{hyperref}
\usepackage{epstopdf}
\usepackage{appendix}

\definecolor{navyblue}{rgb}{0,0.08,0.45}

\def\Dslash{\raise.15ex\hbox{/}\kern-.7em D}
\def\Pslash{\raise.15ex\hbox{/}\kern-.7em P}

 \newcommand{\threehalf}{{\frac{3}{2}}}

\newcommand{\beq}{\begin{equation}}
\newcommand{\enq}{\end{equation}}
\newcommand{\beqa}{\begin{eqnarray}}
\newcommand{\beqast}{\begin{eqnarray*}}
\newcommand{\enqa}{\end{eqnarray}}
\newcommand{\enqast}{\end{eqnarray*}}
\newcommand{\beml}{\begin{multline}}
\newcommand{\enml}{\end{multline}}
\newcommand{\nn}{\nonumber}
\newcommand{\req}[1]{(\ref{#1})}

\newcommand{\pa}{\partial}

\newcommand{\bec}{\begin{center}}
\newcommand{\enc}{\end{center}}
\newcommand{\beqo}{\begin{quote}}
\newcommand{\enqo}{\end{quote}}

\newcommand{\half}{{\textstyle{\frac{1}{2}}}}

\newcommand{\ze}{\zeta}

\newcommand{\la}{\lambda}

\newcommand{\vp}{\varphi}

\begin{document}

\begin{flushright}
{
\small
SLAC--PUB--15892\\
\date{today}}
\end{flushright}

\vspace{60pt}

\centerline{\Large \bf Hadron Spectroscopy and Dynamics}

\vspace{10pt}

\centerline{\Large \bf from Light-Front Holography and Conformal Symmetry}

\vspace{20pt}

\centerline{{
Guy F. de T\'eramond,$^{a}$ 
\footnote{E-mail: \href{mailto:gdt@asterix.crnet.cr}{gdt@asterix.crnet.cr}}
Stanley J. Brodsky,$^{b}$ 
\footnote{E-mail: \href{mailto:sjbth@slac.stanford.edu}{sjbth@slac.stanford.edu}}
and
Hans G\"unter Dosch$^{c}$ 
\footnote{E-mail: \href{mailto:gdt@asterix.crnet.cr}{dosch@thphys.uni-heidelberg.de}}
}}

\vspace{30pt}

{\centerline {$^{a}${\it Universidad de Costa Rica, San Jos\'e, Costa Rica}}

\vspace{4pt}

{\centerline {$^{b}${\it SLAC National Accelerator Laboratory, 
Stanford University, Stanford, CA 94309, USA}}

\vspace{4pt}

{\centerline {$^{c}${\it Institut f\"ur Theoretische Physik, Philosophenweg 16, D-6900 Heidelberg, Germany}}

 \vspace{60pt}

\begin{abstract}

To  a first semiclassical approximation one can reduce the multi-parton light-front problem in QCD to an effective one-dimensional quantum field theory, which encodes the fundamental conformal symmetry of the classical QCD Lagrangian. This procedure  leads to a relativistic light-front wave equation for arbitrary spin which incorporates 
essential spectroscopic and non-perturbative dynamical features of hadron physics.  The mass scale for confinement and higher dimensional holographic mapping to AdS space are also emergent properties of this framework.

\end{abstract}

\newpage

\section{Introduction}
\label{intro}

Recent analytical insights into the nonperturbative nature of the confining interaction in QCD follow from the remarkable holographic correspondence between the equations of motion in  AdS space  and the light-front (LF) Hamiltonian equations of motion for relativistic light hadron bound-states in physical space-time~\cite{deTeramond:2008ht}.   
In fact, the mapping of the equations of motion~\cite{deTeramond:2008ht}  and the matching of the  electromagnetic~\cite{Brodsky:2006uqa, Brodsky:2007hb} and gravitational~\cite{Brodsky:2008pf} form factors in AdS space~\cite{Polchinski:2002jw, Abidin:2008ku} with the corresponding expressions  derived from LF quantization in physical space time is the central feature of the LF holographic approach to hadronic physics.  This  approach allows us to establish a precise relation between wave functions in AdS space and the LF wavefunctions (LFWFs) describing the internal structure of hadrons.  However the actual form of the effective potential has remained unknown until very recently and is thus model-dependent.

It was been realized very recently~\cite{Brodsky:2013ar} that the form of the effective LF confining potential can be obtained from the framework introduced in a remarkable paper by V. de Alfaro, S. Fubini and G. Furlan (dAFF)~\cite{deAlfaro:1976je}. It was shown by dAFF   that in the Schr\"odinger representation, a scale can appear in the Hamiltonian operator while retaining the conformal invariance of the action~\cite{deAlfaro:1976je}. This remarkable result is based on the isomorphism of the algebra of the one-dimensional conformal group Conf$(R^1)$ to the algebra of generators of the group SO(2,1). One of the generators of this group, the rotation in the 2-dimensional space, is compact and has therefore a discrete spectrum with normalizable eigenfunctions. As a result, the form of the evolution operator  is fixed and includes a confining harmonic oscillator potential, and the time variable has a finite range.  In fact, it was shown in Ref. \cite{Brodsky:2013ar} that there exists a remarkable  holographic connection between the one-dimensional semiclassical approximation to light-front dynamics with gravity in a higher dimensional AdS space, and the constraints imposed by the invariance properties under the full conformal group in one dimension Conf$(R^1)$. Other approaches to emergent holography are discussed in~\cite{Koch:2010cy, Glazek:2013jba, Qi:2013caa, Dietrich:2013kza}.

\section{Light-Front Holography and Conformal Quantum \\ Mechanics}

For a hadron with four-momentum $P^\mu$, the generators $P^\mu = (P^-, P^+,  \vec{P}_\perp)$, $P^\pm = P^0 \pm P^3$,  are constructed canonically from the QCD Lagrangian by quantizing the system on the light-front at fixed LF time $x^+$, $x^\pm = x^0 \pm x^3$~\cite{Brodsky:1997de}. The LF Hamiltonian $P^-$ generates the LF time evolution 
$P^- \vert \phi \rangle = i \frac{\pa}{\pa x^+} \vert \phi \rangle$,  whereas the LF longitudinal $P^+$ and transverse momentum $\vec P_\perp$ are kinematical generators.
In the limit of zero quark masses the longitudinal modes decouple  from the  invariant  LF Hamiltonian  equation  $H_{LF} \vert \phi \rangle  =  M^2 \vert \phi \rangle$
with  $H_{LF} = P_\mu P^\mu  =  P^- P^+ -  \vec{P}_\perp^2$. We obtain the wave equation~\cite{deTeramond:2008ht}
\beq \label{LFWE}
\left(-\frac{d^2}{d\ze^2}
- \frac{1 - 4L^2}{4\ze^2} + U\left(\ze, J\right) \right)
\phi_{n,J,L} = M^2 \phi_{n,J,L},
\enq
a relativistic single-variable  LF Schr\"odinger equation, where  the effective potential $U$ acts on the valence sector of the theory.
The effective potential follows from the systematic expression of the higher Fock components as functionals of the lower ones~\cite{Pauli:1998tf}.
The variable $z$ of AdS space is identified with the LF   boost-invariant transverse-impact variable $\zeta$~\cite{Brodsky:2006uqa}, 
thus giving the holographic variable a precise definition in LF QCD~\cite{deTeramond:2008ht, Brodsky:2006uqa}.
 For a two-parton bound state $\zeta^2 = x(1-x) b^{\,2}_\perp$,
where $x$ is the longitudinal momentum fraction and $ b_\perp$ is  the transverse-impact distance between the quark and antiquark.

Recently we have derived wave equations for hadrons with arbitrary spin starting from an effective action in  AdS space~\cite{deTeramond:2013it}.    An essential element is the mapping of the higher-dimensional equations  to the LF Hamiltonian equation  found in Ref.~\cite {deTeramond:2008ht}.  This procedure allows a clear distinction between the kinematical and dynamical aspects of the LF holographic approach to hadron physics.  Accordingly, the non-trivial geometry of pure AdS space encodes the kinematics,  and the additional deformations of AdS encode the dynamics, including confinement~\cite{deTeramond:2013it}, and determine the form of the LF effective potential from the precise holographic mapping to light-front physics~\cite{deTeramond:2008ht, deTeramond:2013it}.  For $d=4$ one finds  from the dilaton-modified AdS action the LF potential~\cite{deTeramond:2013it, deTeramond:2010ge}
\beq \label{U}
U(\ze, J) = \frac{1}{2}\vp''(\ze) +\frac{1}{4} \vp'(\ze)^2  + \frac{2J - 3}{2 \zeta} \vp'(\ze) ,
\enq
provided that the product of the AdS mass $m$ and the  AdS curvature radius $R$ are related to the total and orbital angular momentum, $J$ and  $L$ respectively, according to $(m  R)^2 = - (2-J)^2 + L^2$.  The critical value  $J=L=0$  corresponds to the lowest possible stable solution, the ground state of the LF Hamiltonian, in agreement with the AdS stability bound   $(m R)^2 \ge - 4$~\cite{Breitenlohner:1982jf}, where $R$ is the AdS radius.

The classical Lagrangian of QCD is, in the limit of massless quarks, invariant under conformal transformations. 
Since we are interested in a semiclassical approximation to the nonperturbative domain of QCD, analogous to the quantum mechanical wave equations  in atomic physics, it is natural to have a closer look at conformal quantum mechanics, a conformal field theory in one dimension, following dAFF~\cite{deAlfaro:1976je}.  While leaving the action invariant, the dAFF procedure  leads to a redefinition of the quantum mechanical evolution operator, and consequently to a redefinition of the  corresponding `time' evolution parameter $\tau$, the range of which is finite. In the Schr\"odingier representation
$$
i \frac{\partial}{\partial \tau} \psi(x,\tau) = H_\tau \Big(x,   - i \frac{d}{d x} \Big) \psi(x,\tau),
$$
the dAFF Hamiltonian $H_\tau$ is given by~\cite{deAlfaro:1976je, Brodsky:2013ar}
\beq \label{Htaux}
H_\tau = \frac{1}{2} u \left(- \frac{d^2}{d x^2} + \frac{g}{x^2}\right)  + \frac{i}{4} v  \left(x \, \frac{d}{d x}+ \frac{d}{d x} \, x  \right) +\frac{1}{2}  w x^2, \\
             \enq
which is at  $\tau = 0$, the superposition of the `free' Hamiltonian $H$, the generator of dilatations $D$ and the generator of special conformal transformations $K$ in one dimension, the generators of Conf$(R^1)$; namely 
$$H_\tau = u H + v D + w K.$$ 
The conformal group Conf$(R^1)$ is locally isomorphic to  SO(2,1),  the Lorentz group in 2+1 dimensions. Since the generators of Conf$(R^1)$ have different dimensions, their relations with the generators of SO(2,1) imply a scale, which here plays a fundamental role, as already conjectured in~\cite{deAlfaro:1976je}.

Comparing the dAFF  Hamiltonian \req{Htaux}  with the light-front wave equation \req{LFWE} and identifying the variable $x$ with the light-front invariant variable $\zeta$,  we have to choose $u=2, \; v=0$ and relate the dimensionless constant $g$ to the LF orbital angular momentum, $g=L^2-1/4$,  in order to reproduce the light-front kinematics. Furthermore  $w = 2 \lambda^2$ fixes the confining light-front  potential to a quadratic $\la^2 \, \zeta^2$ dependence.  The choice of the dilaton profile $\varphi(z) = \lambda z^2$ introduced in~\cite{Karch:2006pv} thus follows  from the requirements of conformal invariance. This specific form for $\varphi(z)$  leads through \req{U} to the effective LF potential 
$$U(\ze, J) =   \la^2 \ze^2 + 2 \la (J - 1),$$ 
and corresponds to a transverse oscillator in the light-front. The term $\la^2 \ze^2$ is determined uniquely by the underlying conformal invariance of classical QCD incorporated in the one-dimensional effective theory, and the constant term  $2 \la (J - 1)$ by the embedding space~\cite{deTeramond:2013it, deTeramond:2010ge}.

\subsection{A Light-Front Holographic Model for Mesons}

From \req{LFWE} one obtains for the effective  LF potential $U(\ze, J) =   \la^2 \ze^2 + 2 \la (J - 1)$  for $\lambda >0$ a mass spectrum for mesons characterized by the total angular momentum $J$, the orbital angular momentum $L$ and orbital excitation $n$ given by
$$
M_{n, J, L}^2 = 4 \la \left(n + \frac{J+L}{2} \right),
$$
an important result also found in Ref. \cite{Gutsche:2011vb}.
This result not only implies linear Regge trajectories, but also a massless pion and the relation between the $\rho$ and $a_1$ mass usually obtained from the 
Weinberg sum rules~\cite{Weinberg:1967kj}. The  model  also predicts hadronic LFWFs which underlie form factors~\cite{deTeramond:2012rt} and other dynamical observables, as well as vector meson electroproduction~\cite{Forshaw:2012im}. 
\begin{figure}[h]
\centering
\includegraphics[width=6.8cm]{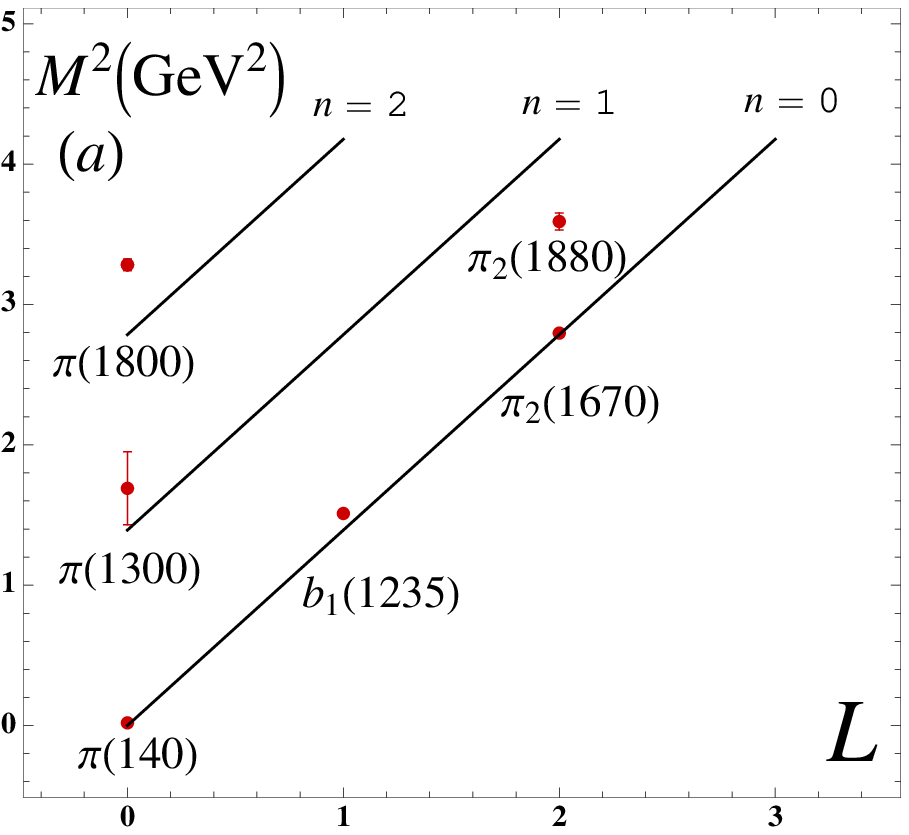}  \hspace{40pt}
\includegraphics[width=6.8cm]{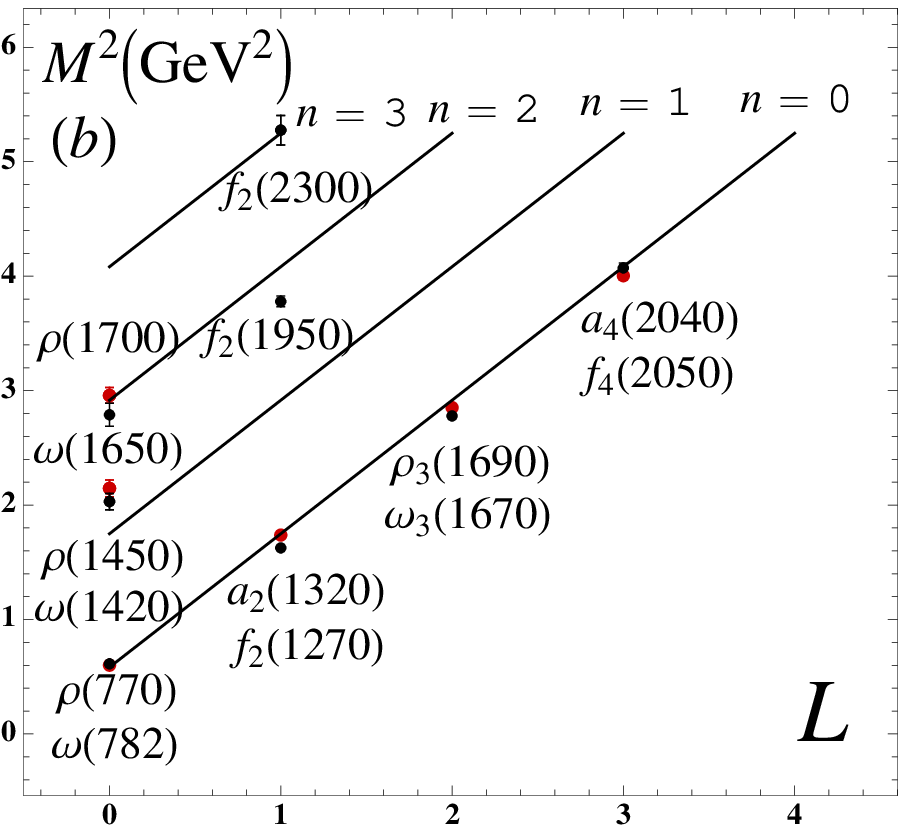}
 \caption{\small $I\!=\!1$ parent and daughter Regge trajectories for the light pseudoscalar mesons (a) with
$\sqrt \la = 0.59$ GeV; and  $I=0$ and $I=1$  light vector-mesons
 (b) with $\sqrt \la = 0.54$ GeV.}
\label{pionspec}
\end{figure} 

The spectral predictions  for the $J = L + S$ light pseudoscalar and vector meson  families are  compared with experimental data in Fig. \ref{pionspec}.
The data are from PDG~\cite{PDG:2012}.   As we will discuss in the next section, a spin-orbit effect is only predicted  for mesons not baryons, as observed in experiment;  it thus becomes a crucial test for any model which aims to describe the systematics of the light hadron spectrum. Using the spectral formula for $M^2$ given above, we find~\cite{deTeramond:2012rt}
$$
 M_{a_2(1320)} >  M_{a_1(1260)} >   M_{a_0(980)}.
$$
The predicted values are 0.76, 1.08 and 1.32 GeV for the masses of the  $a_0(980)$, $a_1(1260)$ and   $a_2(1320)$, compared with the experimental values 0.98, 1.23 and 1.32  GeV respectively. The prediction for the mass of the $L=1$, $n=1$  state $a_0(1450)$ is  1.53 GeV, compared with the observed value 1.47 GeV. The LF holographic model with $\la>0$ accounts for the mass pattern observed in the radial and orbital excitations of the light mesons, as well as for the triplet splitting for the $L=1$, $J = 0, 1, 2$, vector meson $a$-states. The slope of the Regge trajectories gives a value $\sqrt \la \simeq 0.5 ~\rm GeV$. The value of $\la$ required for describing the pseudoscalar sector is slightly higher that the value of $\la$ extracted from the vector sector.  The prediction for  the observed spin-orbit splitting for the $L=1$  $a$-vector mesons  is overestimated by the model.  Note that the solution for $\lambda < 0$ leads to a pion mass heavier than the $\rho$ meson in clear disagreement with observations.

\subsection{A Light-Front Holographic Model for Baryons}

The analytical exploration of the baryon spectrum using light-front holographic ideas is not as simple or as well understood as the meson case. However, as we shall discuss is this section,   even a relatively simple approach provides a framework for a useful description of the baryon spectrum which gives important insights into its systematics.  In a chiral spinor component representation,  the light-front wave equations for baryons  are given by the coupled linear differential equations~\cite{deTeramond:2013it}
\begin{eqnarray} \label{LFDE}  
- \frac{d}{d\zeta} \psi_-  - \frac{\nu+\half}{\zeta}\psi_-  -  V(\zeta) \psi_-  &=& M \psi_+ ,  \nn \\
 \frac{d}{d\zeta} \psi_+ - \frac{\nu+\half}{\zeta}\psi_+  - V(\zeta) \psi_+  &=&  M \psi_- , 
\end{eqnarray}
where $\nu$ is given in terms of the mass appearing in the Dirac equation in AdS space: $ \nu = \vert \mu R \vert - \half $.  One can also show that  the effective potential $V$ is  $J$-independent~\cite{deTeramond:2013it, Gutsche:2011vb}. This is a remarkable result, since it implies that independently of the specific form of the potential, the value of the baryon masses along a given Regge trajectory depends only on the LF orbital angular momentum $L$, and thus, in contrast with the vector mesons, there is no spin-orbit coupling, in agreement with the observed near-degeneracy in the baryon spectrum.

We choose  an effective  linear confining potential $V = \lambda \,  \zeta$ which also leads to linear Regge trajectories in both the orbital and radial quantum numbers for baryon excited states. The linear potential also leads to the LF oscillator form  $\lambda ^2 \zeta^2$ in the second order version of Eqs. \ref{LFDE}.  For $\la > 0$ we find the result
$$
M^2 = 4 \, \lambda  \left(  n +  \nu + 1 \right).
$$
For $\la <0$  no solution is possible~\cite{deTeramond:2013it}.  To determine the internal spin, internal orbital angular momentum and radial quantum number assignment of the $N$ and $\Delta$ excitation spectrum from the total angular momentum-parity PDG assignment, it is  convenient to use the conventional ${\rm SU(6)} \supset {\rm SU(3)}_{flavor} \times {\rm SU(2)}_{spin}$ multiplet structure.  
\begin{figure}[ht]
\centering
\includegraphics[width=6.8cm]{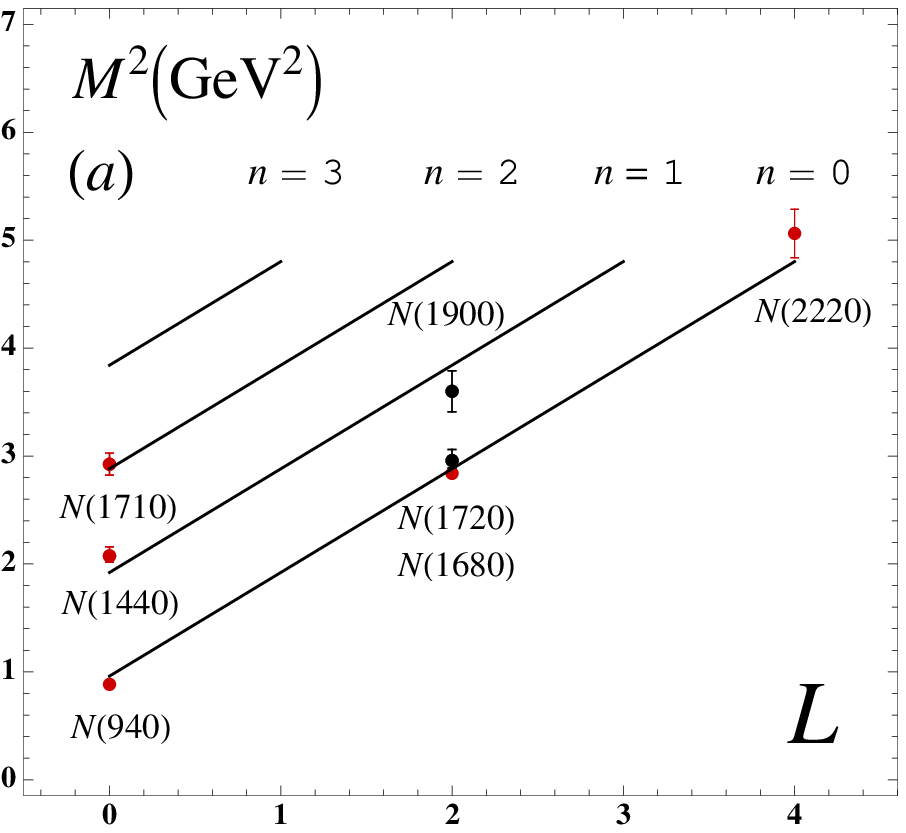}   \hspace{40pt}
\includegraphics[width=6.8cm]{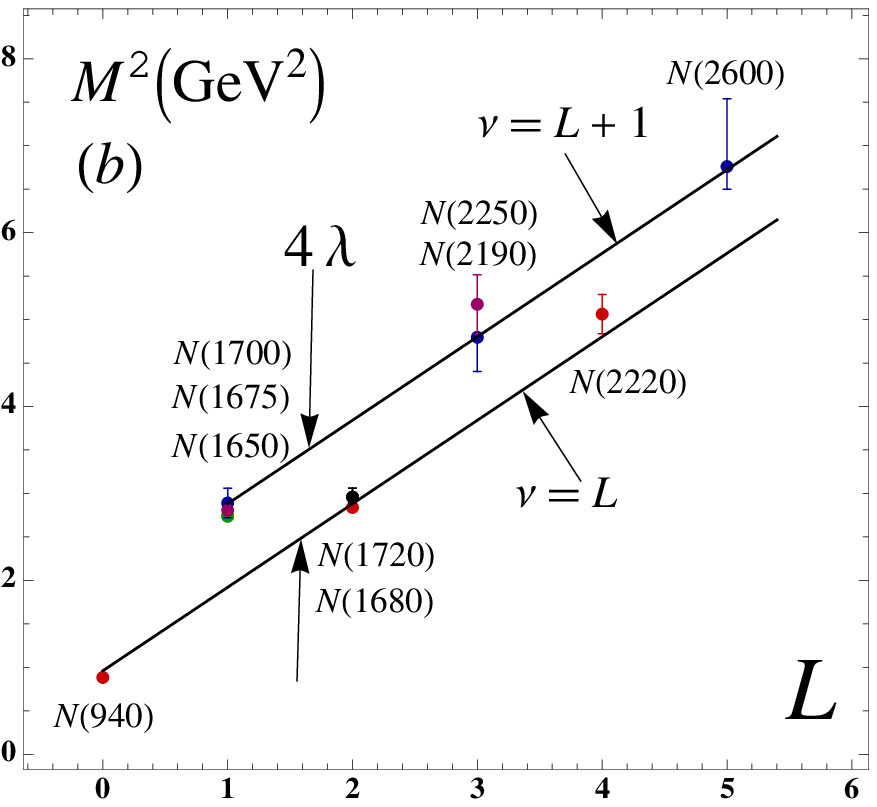}  

 \vspace{30pt}
 
 \includegraphics[width=6.8cm]{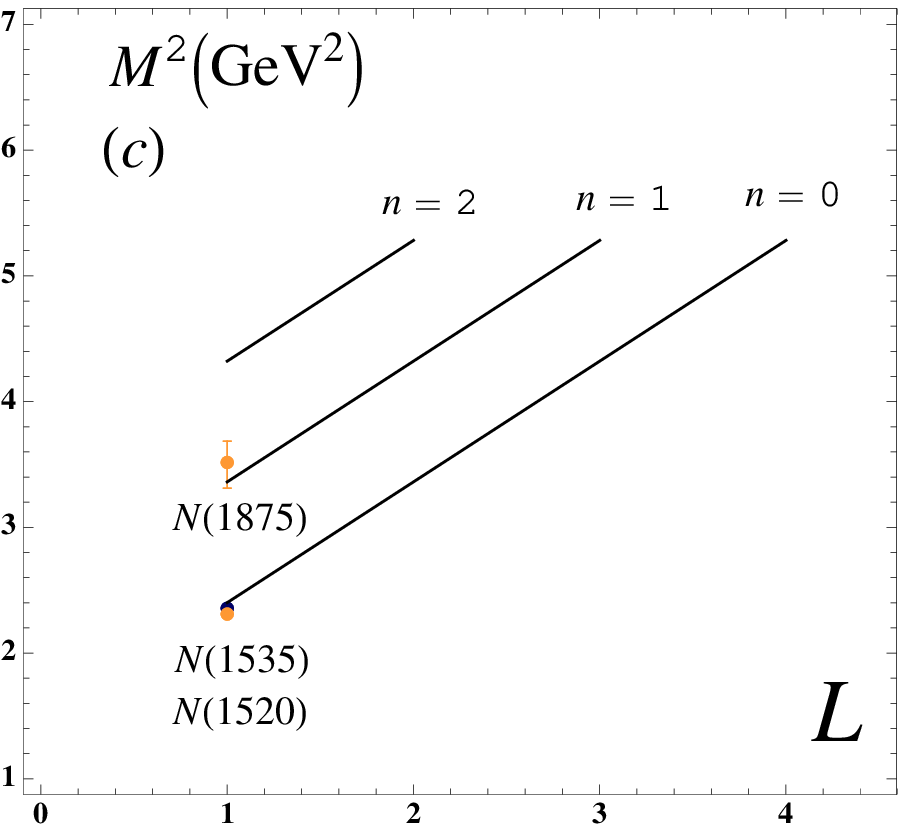}  \hspace{40pt}
\includegraphics[width=6.8cm]{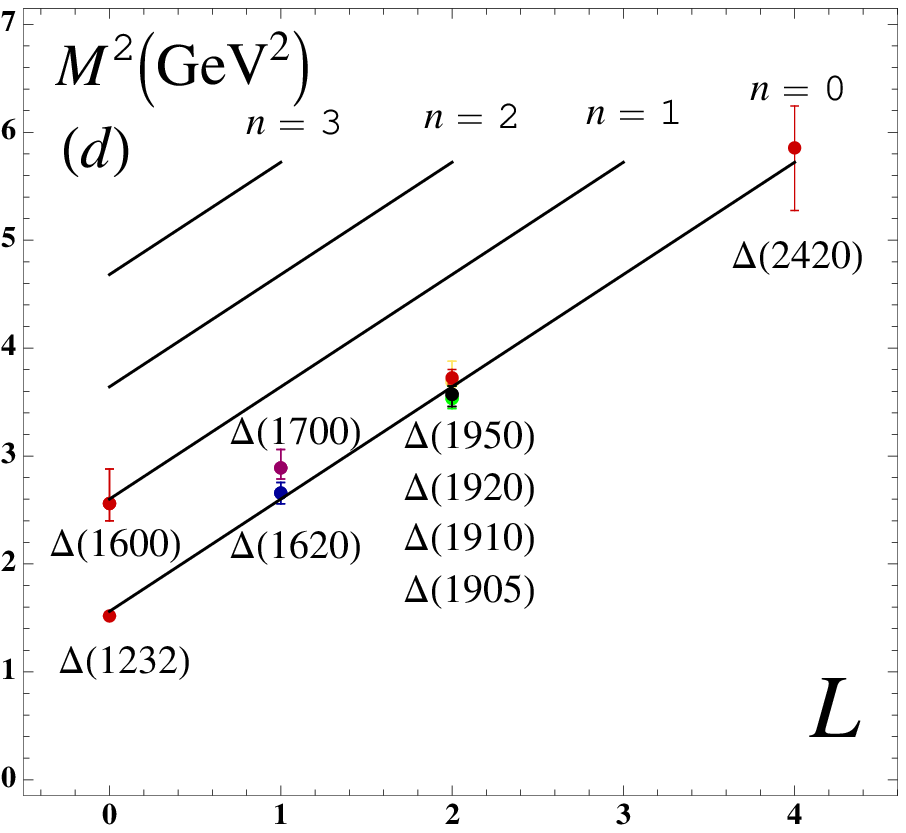}
 \caption{\small Orbital and radial baryon excitation spectrum.  Positive-parity spin-$\half$ nucleons (a) and spectrum gap between the negative-parity spin-$\threehalf$ and the positive-parity spin-$\half$ nucleons families (b). Minus parity $N$ ({c}) and plus and minus parity $\Delta$  families (d), for $\sqrt{\la} = 0.49$ GeV (nucleons) and  0.51 GeV (Deltas).}
 \label{baryonspec}
\end{figure} 

The lowest possible stable state,  the proton,  corresponds to $n=0$ and $\nu = 0$. This fixes the scale $\sqrt \lambda \simeq 0.5$ GeV. The resulting predictions for  the positive-parity spin-$\half$ nucleons are shown in  Fig. \ref{baryonspec} (a) for  the parent Regge trajectory for $n =0$ and  $\nu = 0, 2, 4, \cdots,  L $, where $L$ is the relative LF angular momentum between the active quark and the spectator cluster. The predictions for the daughter trajectories for $n=1$, $n = 2, \cdots $ are also shown in this figure. Only confirmed PDG \cite{PDG:2012} states are shown. The Roper state $N(1440)$ and the $N(1710)$ are well accounted for  as the first and second radial excited states of the proton. The newly identified state, the $N(1900)$~\cite{PDG:2012} is depicted here as the first radial excitation of the $N(1720)$. The model is successful in explaining the parity degeneracy observed in the light baryon spectrum, such as the $L=2$, $N(1680) - N(1720)$  pair in Fig. \ref{baryonspec} (a).  In Fig.  \ref{baryonspec} (b) we compare the positive parity  spin-$\half$ parent nucleon trajectory with the negative parity spin-$\threehalf$ nucleon trajectory.  It is remarkable that the gap scale $4 \la$ determines not only the slope of  the trajectories, but also the gap in the spectrum  between the plus-parity spin-$\half$  and the minus-parity spin-$\threehalf$ nucleon families,  as indicated by arrows in this figure. This means the respective assignment $\nu = L$ and $\nu = L+1$ for the lower and upper trajectories in Fig. \ref{baryonspec} (b). We  also note that the degeneracy of states with the same orbital quantum number $L$ is also well described, as for example the degeneracy of the $L=1$ states  $N(1650)$, $N(1675)$ and $N(1700)$ in Fig. \ref{baryonspec} (b).

We have also to take into account baryons  with negative parity and internal spin $S = \half$, as well as baryon states with positive parity and internal spin $S = \threehalf$ such as the $\Delta(1232)$. Those states are well described by the assignment $\nu = L + \half$. This means, for example, that  $M^{2 \,(+)}_{n, L, S = \frac{3}{2}} = M^{2 \,(-)}_{n, L, S =  \frac{1}{2}}$ and consequently the positive and negative-parity $\Delta$ states lie in the same trajectory consistent with the experimental results, as depicted in Fig. \ref{baryonspec} (d). The newly found state, the $N(1875)$ \cite{PDG:2012}, depicted in Fig. \ref{baryonspec} ({c}) is well described as the first radial excitation of the $N(1520)$, and the near degeneracy of the $N(1520)$ and $N(1535)$ is also well accounted. Likewise, the  $\Delta(1600)$ corresponds to the first radial excitation of the $\Delta(1232)$ as shown in Fig. \ref{baryonspec} (d).  The model explains the important degeneracy of  the $L=2$, $\Delta(1905), $ $ \Delta(1910),$ $\Delta(1920), $ $ \Delta(1950)$ states which are degenerate within error bars.  Our results for the $\Delta$ states agree with those of Ref.~\cite{Forkel:2007cm}. ``Chiral partners" such as the $N(1535)$ and the $N(940)$ with the same total angular momentum $J = \half$, but with different orbital angular momentum are non-degenerate from the onset.  
 To recapitulate, the parameter $\nu$  has the internal spin $S$ and parity $P$ assignment  given in  the table below.
\begin{table}[ht] \label{nuT}
\small
\centering 
\caption{\small Orbital assignment for baryon trajectories  according to parity and internal spin.}
\vspace{5pt}
\resizebox{7cm}{!} {
\begin{tabular}{ l | c r } 
 & $S = \half$ & \,$S = \threehalf$ \ ~~~~~\\ [1.0ex]
 \hline
P = + & $\nu = L$ &  $ \nu = L + \half$ \\ [1.0ex]
 P = \ -  &  $\nu = L + \half$ &  $\nu = L+1$ \\ [0.6ex]
 \hline
\end{tabular}
\label{asig}
}
\end{table}

The assignment     $\nu = L$ for the lowest trajectory, the proton trajectory,  is straightforward and follows from the mapping of AdS to light-front physics. The assignment for other spin and parity baryons states  given in Table \ref{asig} is phenomenological. It is expected that further analysis of the different quark, or quark--diquark, configurations and symmetries  of the baryon wave function will indeed explain the actual assignment in Table \ref{asig}, which  successfully describes the full light baryon orbital and radial excitation spectrum, and in particular the gap between trajectories with different parity and internal spin.  The holographic variable $\zeta$ has a cluster decomposition and thus labels a system of $n$-partons as an active quark plus a system of $n-1$ spectators~\cite{Brodsky:2006uqa}. From this perspective, a baryon with $n=3$ looks in light-front holography as a quark--diquark system.

\section{Conclusions}

We have followed the remarkable results of De Alfaro, Fubini and Furlan~\cite{deAlfaro:1976je}  which, combined with light-front holographic QCD~\cite{deTeramond:2008ht, Brodsky:2013ar}, give important insights into the  QCD confining mechanism. It turns out that it is possible to introduce a scale by modifying the variable of dynamical evolution and  nonetheless the underlying action stays conformally invariant. This procedure determines uniquely the form of the LF effective potential, corresponding to a quadratic dilaton  modification  of AdS space to include confinement.   We obtain an effective  one-dimensional quantum field theory which retains the fundamental conformal symmetry of the classical QCD Lagrangian in the limit of massless quarks. As a result the mass scale for confinement and  holographic mapping to AdS space are  emergent properties. The group theoretical arguments based on the underlying conformality fix the quadratic form  of the effective dilaton profile and thus the corresponding form of the confinement potential of the LF QCD Hamiltonian. The result is a relativistic LF wave equation for bound states which leads to a remarkable description of the spectroscopy and dynamics of the light mesons and baryons~\cite{deTeramond:2012rt}.

\section*{Acknowledgements}

Invited plenary talk, presented by GdT at the 13$^{th}$ International Conference on Meson-Nucleon Physics and the Structure of the Nucleon (MENU 2013),  Rome, September 30 -- October 4, 2013. GdT wants to thank the organizers for the splendid hospitality in Rome.

\end{document}